\documentclass[pre,twocolumn,floatfix,showpacs,amsmath]{revtex4}
\usepackage{graphicx}
\usepackage{graphics}
\newcommand{\be}{\begin{equation}}
\newcommand{\ee}{\end{equation}}
\newcommand{\bea}{\begin{eqnarray}}
\newcommand{\eea}{\end{eqnarray}}

\begin{document}

\title{Random Walkers with Shrinking Steps in d-Dimensions \\and \\Their Long Term Memory}

\author{Tongu\c{c} Rador}
\email[]{tonguc.rador@boun.edu.tr}
\affiliation{Bo\~{g}azi\c{c}i University \\ Department of Physics \\ Bebek 34342, \.Istanbul, Turkey}

\date{\today}

\begin{abstract}
We study, in d-dimensions, the random walker with geometrically shrinking step sizes at each hop. We emphasize the integrated quantities such as expectation values, cumulants and moments rather than a direct study of the probability distribution. We develop a $1/d$ expansion technique and study various correlations  of the first step to the position as time goes to infinity. We also show and substantiate with a study of the cumulants that to order $1/d$ the system admits a continuum counterpart equation which can be obtained with a generalization of the ordinary technique to obtain the continuum limit. We also advocate that this continuum counterpart equation, which is nothing but the ordinary diffusion equation  with a diffusion constant decaying exponentially in continuous time, captures all the qualitative aspects of the discrete system and is often a good starting point for quantitative approximations. 
\end{abstract}

\pacs{02.50.-r, 05.40-a}

\maketitle

\section{Introduction}

An interesting extension of the ordinary random walk problem is a model in which the walker takes steps with decreasing size at each hop. There has been some interest among the physics community in the study of a particular system of this type during the past years namely the case for which the step lengths are given by $\lambda^{n}$, where $n$ is the time step and $\lambda<1$. The system and its generalities are exposed in \cite{art1} and \cite{art2} for a one dimensional walker and its first passage properties are studied in \cite{art3}, again for the one dimensional case. In this work we extend the study to d-dimensions and try to lay out important aspects within the room available. 

Unfortunately there is no immediate exact solution to the probability distribution for such a walker and consequently one is almost forced to invoke numerical simulations. To that end we tried to shy away from listing many probability density plots for various dimensions and for various values of $\lambda$. Instead we focused on quantities which do not necessarily require a direct knowledge of the probability density, such as moments, cumulants and other interesting expectation values. 

In implementing the choice for hopping directions we  use random unit vectors instead of picking orthogonal directions. This approach, as discussed in the text, keeps rotational symmetry for the probability distribution at all times. Using orthogonal directions for hopping vectors could also be interesting in of itself for other purposes but we do not pursue them here.

One of the important aspects of the system at hand is a generalization of the shift symmetries of the ordinary random walker; namely if one shifts time with one step one also has to shift the position variable by one step and sum among all possible directions. This symmetry is altered due to the geometric change in the step lengths, the translation of the position variable is now accompanied with a scale change of distances.
We use this property to show that  a recursive technique is available for an exact computation of the moments of the distribution, even when time becomes infinity where such an approach is not available for the ordinary random walker. This scaling property married with a $1/d$ expansion approach proves to be very useful in computing various other quantities such as fractional power moments.
 
We use the $1/d$ expansion technique to study lower order cumulants of the probability distribution and contrast them to those of a gaussian. We find that to order $1/d$ the probability distribution has vanishing cumulants except for the variance which is exactly what one finds for a gaussian. However the properly normalized variances do not agree and this shows that the distribution is not exactly gaussian but very close to it. To that end we show that the $1/d$ approach allows for a continuum counterpart system -not a naive continuum limit- and the mentioned discrepancy is tied to the fact that for $\lambda<1$  there is no continuum limit. Nevertheless we show that the continuum counterpart system is one for which the  diffusion parameter decays exponentially with time.

Since $\lambda<1$, the walker eventually comes to a halt in the infinite time limit and this means that there will be a non-zero probability density defined in a finite support characterized by the maximum distance the walker can take. This is in contrast to the
ordinary random walker ($\lambda=1$) where the density vanishes in this limit. Due to this effect the choice for the first step (of length $1$) is strongly correlated to the final position of the walker. We present the study of two such correlations. The first one is the expectation value of the cosine of the angle between the first step and the position after infinitely many steps are taken. We present the numerical simulations and compare it to the $1/d$ expansion method developed in the text. The second long term memory effect we study is the probability, in the infinite time limit, for the walker to be along the $+\hat{z}$ direction, given that after the first step the walker is at $+\hat{z}$. This observable does not have a $1/d$ expansion --it depends on $\sqrt{d}$--, however using the continuum counterpart model as a starting point we find a phenomenological function which fits the numerical data rather well for all $d$ and $\lambda$.

\section{The Model}\label{sec2}
Consider a random walker in d-dimensions for which the step sizes decrease exponentially with each hop. Then
if the walker starts at the origin, the random variable for its position is simply given by

\be
\vec{r}_{N}=\sum_{0}^{N} \lambda^{n} \hat{r}_n \;\; .
\ee

\noindent Here $\hat{r}_{n}$ are random unit vectors obeying the following statistical rules

\begin{subequations}
\bea
\left<\hat{r}_{n}\right>&=&0 \;\;, \\
\left<\hat{r}_{n}\cdot\hat{r}_{m}\right>&=&\delta_{nm}\;\;.
\eea
\end{subequations}

\noindent Consequently we have (in any dimensions) the following,

\begin{subequations}
\bea
\left<\vec{r}_\infty\right>&=&0\;\;,\\
\left<r^{2}_\infty\right>&=&\frac{1}{1-\lambda^{2}}\;\;,\\
r_{\infty}^{\rm{max}}&=&\frac{1}{1-\lambda}\;\;.
\eea
\end{subequations}

An interesting property of the walker is that for $\lambda<1/2$ the walker will fall out of range of the origin once it takes the first step. In this case the walker will be confined in a circular band with limits

\be
\frac{1-2\lambda}{1-\lambda}\;\leq\;r\;\leq\;\frac{1}{1-\lambda}\;\;.
\ee

The distribution function for $\vec{r}_{N}$ can also be expressed easily if we make use of the conservation of probability: {\em if the walker is somewhere at step $N+1$ then it must be at a
neighboring point at the previous step}. We therefor get

\be\label{eq:pdf}
P(\vec{r},N+1)=\frac{1}{\mathcal{N}}\;\; \sum_{\hat{r}_{N}} P(\vec{r}-\lambda^{N}\hat{r}_{N},N)\;\;.
\ee

\noindent With the normalization factor given by

\be
\mathcal{N}=\sum_{\hat{r}_{N}}\;1\;\;.
\ee

From our notation it is immediate that the probabilities to pick the directions are evenly distributed. This equation is nothing but a total sum over the possible neighbors normalized by the total
number of neighbors. If one uses  random unit vectors to pick hopping directions, 
$\mathcal{N}$ simply stands for the integral over the angular measure in a d-dimensional space.

From (\ref{eq:pdf}) it is clear that as $N\to\infty$ the probability distribution  approaches a time independent function which we will denote as $P_{\lambda}(\vec{r})$.

\subsection{Rotational Invariance}

As far as rotational invariance is concerned, there is an important property of the walker we have presented. To expose this let us consider the model in 2-dimensions: if one uses orthogonal directions to pick
the equal weight random hopping vectors, the distribution function $P_{\lambda}(\vec{r})$ will not necessarily have rotational invariance. This effect may be understood clearer if one considers $\lambda=0$ where it
is most pronounced. Since for $\lambda=0$ the walker stops after the first step the probability distribution is,

\be
\frac{1}{4}\left[\delta(\vec{r}-\hat{x})+\delta(\vec{r}+\hat{x})+\delta(\vec{r}-\hat{y})+\delta(\vec{r}+\hat{y})\right]\nonumber\;\;.
\ee

\noindent This has no continuous rotational invariance, only a discrete one. This effect should remain as 
we increase $\lambda$. In fact one can hope to gain rotational invariance in the infinite time limit only as $\lambda\to 1$. The slowing down of the walker breaks the rotational invariance if one insists on picking the random vectors along the orthogonal directions.

As described in \cite{art1} this peculiarity is related to the fact that the support of the distribution function is not a regular lattice: it is a collection of points that evolves in time in a rather complicated way. In studying the {\em ordinary} random walker introducing a lattice makes sense since one knows that at each time step the walker resides on the backbone lattice and that in the infinite time limit rotational invariance is restored. However in the model presented here choosing lattice schemes for the choice of random hopping vectors will not recover rotational invariance in full as time goes to infinity. Using random unit vectors to pick a hopping direction has {\em a priori} rotational symmetry. This is the reason why we have implemented the model with random unit vectors in the beginning of this section. On the other hand depending on the particular natural system where this model can be applicable, choosing a lattice scheme can be justified and could be of interest in of itself. In this work we will be working with the continuous random unit vector scheme. 

\section{General Properties of $P_{\lambda}$}

In this section we try to expose general properties of the probability distribution. In subsection A we study the Fourrier transform technique and present the solution for arbitrary $d$. In the next part we discuss the scaling property of the distribution and show that it allows for a recursive method to compute moments of the distribution. Subsection C is reserved for the development of a $1/d$ expansion technique which we use in part D to study the variance, skewness and kurtosis of the distribution and contrast it to a gaussian. We also show in the last subsection that a  continuum counterpart system is available as the first term in a $1/d$ expansion of the master equation (\ref{eq:pdf}) . The continuum counterpart system is not a naive continuum limit and has a diffusion constant that decays exponentially with time.

\subsection{Fourrier Space Solution}

One can try to make an explicit use of (\ref{eq:pdf}) by introducing the Fourier transform $\tilde{P}_{\lambda}$ of $P_{\lambda}$ which which will satisfy the following equation,

\be{\label{eq:fourrier}}
\tilde{P}_{\lambda}(k,N+1)=\mathcal{N}\;\tilde{P}_{\lambda}(k,N)\;\sum_{\hat{r}_{N}}\;\exp\left[-i\lambda^{N}\vec{k}\cdot\hat{r}_{N}\right].
\ee

\noindent In view of rotational invariance we can apply the convention of  picking $\vec{k}$ along the $\hat{z}$ direction in a d-dimensional space. This will result in the following, 

\be{\label{eq:fourrier2}}
\tilde{P}(k,N+1)=\tilde{P}(k,N)\int \frac{d\omega_{d}(x)}{\omega_{d}} \exp\left[-i\lambda^{N} k x\right].
\ee

\noindent Here $x$ is the cosine of the angle between $\hat{r}_{N}$ and the $\hat{z}$ direction. The measure
$d\omega_{d}(x)$ is 

\be{\label{eq:measx}}
d\omega_{d}(x)\equiv g_{d}(x)dx=\left(1-x^{2}\right)^{\frac{d-3}{2}}\;dx\;,
\ee

\noindent the region of integration for $x$ is $[-1,1]$, and $\omega_{d}$ is the integrated measure. The formalism is valid for all $d$ though  one has to exercise attention for $d=2$ and $d=1$. For $d=2$ we have $x=cos(\theta)$ and $d\omega_{2}=d\theta$ with $\theta\in\{0,2\pi\}$, the integrated measure is $2\pi$. For $d=1$ there is no integral,
only a sum over the discrete values of $x=\pm1$, so the normalization factor becomes $2$ 

The solution to (\ref{eq:fourrier2}) is easily obtained remembering that if the walker starts at the origin we
have $\tilde{P}_{\lambda}(k,0)=1$. Then by iteration we have the following,

\be{\label{eq:foursol}}
\tilde{P}_{\lambda}(k,N)=\prod_{n=0}^{N}\;\mathcal{G}(d,\lambda^{n}k)\;\;,
\ee

\noindent with 

\bea
\mathcal{G}(d,k)&=&\frac{1}{\sum_{\hat{r}}1}\;\sum_{\hat{r}} \exp(i\vec{k}\cdot\hat{r})\nonumber\;\;,\\
&=&\int \frac{d\omega_{d}(x)}{\omega_{d}}\;\cos(kx)\;.
\eea

\noindent Where we again have made use of the convention of picking $\vec{k}$ along $\hat{z}$ direction and in passing to the cosine function we have used the evenness of $\tilde{P}_{\lambda}(k)$. This results in the following list,

\begin{subequations}
\bea
\mathcal{G}(1,k)&=&\cos(k)\;,\\
\mathcal{G}(2,k)&=&\rm{J}_{0}(k)\;,\\
\mathcal{G}(3,k)&=&j_{0}(k)\;,\\
\mathcal{G}(4,k)&=&2\rm{J}_{1}(k)/k\;,\\
\mathcal{G}(5,k)&=&3j_{1}(k)/k\;,\\
\mathcal{G}(6,k)&=&8\rm{J}_{2}(k)/k^{2}\;,\\
\mathcal{G}(7,k)&=&15j_{2}(k)/k^{2}\;,\\
\vdots &=& \vdots \nonumber
\eea
\end{subequations}

\noindent Here $\rm{J}_{i}$ and $j_{i}$ are i'th Bessel and spherical Bessel functions respectively. 
This table amounts to the following expressions  for general $d>1$

\bea
\mathcal{G}(2n,k)=(2n-2)!!\;\;\frac{\rm{J}_{n-1}(k)}{k^{n-1}}\;\;,\\
\mathcal{G}(2n+1,k)=(2n-1)!!\;\;\frac{j_{n-1}(k)}{k^{n-1}}\;\;.
\eea

With these constructs one has access to the usual machinery of the Fourrier technique. For example the moments
of the distribution function in the position space will be given by coefficients of $\tilde{P}(k,N)$ in an expansion in $k$. There is however a more useful albeit indirect method which has its roots in the scaling property of the distribution.

\subsection{Scaling Property of $P_{\lambda}(r)$ and $\tilde{P}_{\lambda}(k)$}

From (\ref{eq:foursol}) it is clear that the distribution in Fourrier space enjoys the following scaling rule
along with time translation

\be{\label{eq:scalefour}}
\tilde{P}_{\lambda}(k,N+1)=\tilde{P}_{\lambda}(\lambda k,N)\;\mathcal{G}(d,k)\;\;,
\ee

\noindent or in the infinite time limit

\be{\label{eq:scalefour2}}
\tilde{P}_{\lambda}(k)=\tilde{P}_{\lambda}(\lambda k)\;\mathcal{G}(d,k)\;\;.
\ee

This suggests connections to various interesting physical systems. 
For example taking the logarithm of (\ref{eq:scalefour2}) we find the following

\be{\label{eq:rad1}}
\psi(k) = h(k)+\psi(\lambda k)\;
\ee

\noindent with $\psi=\ln \tilde{P}_{\lambda}(k)$ and $h=-\ln\mathcal{G}(d,k)$. 

In form (\ref{eq:rad1}) is exactly like the free energy per bond acquired by an exact decimation of a thermodynamical system on a lattice with hierarchy \cite{art4}. If the
system admits one relevant critical parameter $t_{c}$, then for $\tau=t-t_{c}$ close to zero the free energy
is (see for example \cite{art5} and \cite{art6})

\be{\label{eq:berker}}
f(\tau)=g(\tau)+b^{-\mathcal{D}}\;f(\lambda\tau)\;.
\ee

\noindent with $b$ being the lattice scaling factor, $\mathcal{D}$ the effective dimension. Shrinking paramater $\lambda$ in this context is the slope of the linearized renormalization group transformation. So the logarithm of the Fourrier transform of the random walker we study is like a thermodynamical system near criticality  with zero effective dimensions, the critical point in this analogy being $k=0$ for our system. This is somewhat expected since as $k\to0$ we probe longer distance behavior of the distribution, in this region since the walker eventually stops at $r^{max}=(1-\lambda)^{-1}$ the distribution has to vanish for larger values. This hints at a slope discontinuity an hence at a possible critical point at $r^{max}$. 

It has also been shown in this context that there are connections \cite{art7} to q-oscillators \cite{art8} and q-differences \cite{art9}. These analogies are interesting in their own rights. But in this paper we aim at a different direction and will not pursue them further. 

This scaling behavior of the distribution depends on the fact that the steps of the walker shrink geometrically with parameter $\lambda$. This is better understood if we remember that position random variable
obeys the following,

\be{\label{eq:scalepos1}}
\vec{r}_{N+1}=\hat{r}_{1}+\lambda\vec{\rho}_{N}\;\;.
\ee

\noindent That is if $\vec{r}_{N+1}$ is a random variable enough to describe the system at time $N+1$ so is
$\vec{\rho}_{N}$ to describe the system at time $N$. Their statistical independence is characterized by the
discrepancy introduced by $\hat{r}_{1}$. This behavior of the position random variables should also manifest itself at the distribution level and we should have

\be{\label{eq:scalepos2}}
P_{\lambda}(\vec{r},N+1)=\left(\frac{1}{\lambda^{d}\sum_{\hat{r}}1}\right)\sum_{\hat{r}}P_{\lambda}\left(\frac{\vec{r}-\hat{r}}{\lambda},N\right)\;.
\ee

\noindent Obviously if $\lambda=1$ (\ref{eq:scalepos2}) and (\ref{eq:pdf}) are identical, for $\lambda\neq1$ they also contain the same information but this is somewhat clouded by the scale transformation.

In this paper we will be studying infinite time limit properties of the distribution. In this limit the scaling
property reads as

\be{\label{eq:scale}}
\vec{r}_{\infty}=\hat{r}_{1}+\lambda\;\vec{\rho}_{\infty}\;.
\ee

\noindent Here $\vec{\rho}_{\infty}$ is a random variable equivalent to  $\vec{r}_{\infty}$ to describe the system, only it is a statistically different one where the difference
is manifested in the random unit vector $\hat{r}_{1}$. This equivalency means that at the level of the probability distributions

\be{\label{eq:scale2}}
P_{\lambda}(\vec{r})=\left(\frac{1}{\lambda^{d}\sum_{\hat{r}}1}\right)\sum_{\hat{r}}P_{\lambda}\left(\frac{\vec{r}-\hat{r}}{\lambda}\right)\;\;.
\ee

\noindent Which follows from (\ref{eq:scalepos2}) as $N\to\infty$. The form in (\ref{eq:scale2}) have been presented in \cite{art2} for the one dimensional case.

\subsection{Scaling Invariance of Expectation Values}

The equivalency represented by (\ref{eq:scale}) also means exact equality if one considers
expectation values, that is one must have for any function $F$

\be
\left<F(\vec{r}_{\infty})\right>=\left<F(\vec{\rho}_{\infty})\right>\;\;.
\ee

We can exploit this scaling behavior as follows. Let us identify the first step $\vec{r}_{1}$ as the $\hat{z}$ direction in a d-dimensional space. In view of \ref{eq:scale} we have the following for any function $F$,

\be
\left<F(\vec{r})\right>={\tilde{\mathcal{N}}}\int d\rho^{d}\;P_{\lambda}(\rho) \;F(\lambda\vec{\rho}+\hat{z})\;\;,
\ee

\noindent which is nothing but a direct consequence of (\ref{eq:scale2}). From know on we are omitting the subscript $\infty$ as it is clear we are working with the distribution function at the infinite time limit. Also at this point we would like to choose our normalization prescription. It is
better for our purposes to work with a distribution normalized along the radial direction only \footnote{The main reason behind this is in a computer simulation with continuous random unit vectors. One actually samples $r^{d-1}P_{\lambda}(r)$ instead of $P_{\lambda}(r)$.}. With this
convention we have the following

\be{\label{eq:esas}}
\left<F(\vec{r})\right>=\int \frac{d\omega_{d}}{\omega_{d}} \;d\rho\;\rho^{d-1}\;P_{\lambda}(\rho) \;F(\lambda\vec{\rho}+\hat{z})
\ee

This final expression could prove more useful than a direct application of the Fourrier transform technique. One immediate consequence of (\ref{eq:esas}) is that it provides a  recursive algorithm for computing the moments of the distribution function. To expose this let us start with the following,

\[
\left<r^{2}\right>=\int \frac{d\omega_{d}}{\omega_{d}} \;d\rho\;\rho^{d-1}\;P_{\lambda}(\rho) \;(1+\lambda^{2} \rho^{2}+2\lambda \rho x)\;\;.
\]

\noindent Since $x$ has zero expectation value under the measure $d\omega_{d}(x)/\omega_{d}$ we have

\[
\left<r^{2}\right>=\int\;d\rho\;\rho^{d-1}\;P_{\lambda}(\rho)\;(1+\lambda^{2}\rho^{2})\;\;,
\]

\noindent but by our normalization of $P_{\lambda}$ this equation is equivalent to the following,

\[
\left<r^{2}\right>=1+\lambda^{2} \left<r^{2}\right>\;\;,
\]

\noindent which in turn yields the correct result,

\[
\left<r^{2}\right>=\frac{1}{1-\lambda^{2}}\;\;.
\]

If one chooses $F=r^{4}$ we get the following equation

\[
(1-\lambda^{4})\left<r^{4}\right>=1+2\lambda^{2}\left<r^{2}\right>\left[1+\frac{2}{d}\right]\;\;,
\]

\noindent where we have used the fact that $x^{2}$ has expectation value $1/d$ under $d\omega_{d}(x)/\omega_{d}$ giving in the correct result\footnote{Coincides with the result found for $d=1$ in \cite{art2}.} 

\be
\left<r^{4}\right>=\frac{1}{(1-\lambda^{2})^{2}}\left[1+\frac{4\lambda^{2}}{1+\lambda^{2}}\frac{1}{d}\right]\;\;.
\ee

Higher moments get more complicated but are still computable with not so much effort. 
Therefor we have a useful recursive algorithm for finding all the even moments of the distribution function. What we need to apply the recursion are the general moments of the measure $d\omega_{d}(x)$,

\begin{subequations}{\label{eq:xmeans}}
\bea
\{x^{p}\}&\equiv&\int \frac{d\omega_{d}(x)}{\omega_{d}}\;x^{p}\;\;,\\
\{x^{2n}\}&=&\frac{\Gamma(d/2)\Gamma(n+1/2)}{\Gamma(1/2)\Gamma(n+d/2)}\;\;,\\
\{x^{2n+1}\}&=&0\;\;.
\eea
\end{subequations}

\subsection{$1/d$ Expansion of Expectation Values}
For more general functions $F$ one could proceed as follows. We have the following direct from our formalism

\be{\label{eq:gene}}
\left<F(r)\right>=\int \frac{d\omega_{d}}{\omega_{d}} d\rho\rho^{d-1}P_{\lambda}(\rho)F(\sqrt{1+\lambda^{2}\rho^{2}+2\lambda\rho x}).
\ee

The complexity of this approach depends on the function $F$. For general dimensions there is no great simplification and one has to directly implement the formalism. Odd dimensions are better in view of the fact that the measure will not have a square root, particularly for $d=3$ one has $d\omega_{d}(x)=dx$, which tends to simplify the approach somewhat. On the other hand after a closer look at the moments of the measure $d\omega_{d}(x)$ in (\ref{eq:xmeans}) one sees the possibility of a $1/d$ expansion since one has

\be{\label{eq:recex}}
\{x^{2(n+1)}\}=\frac{2n+1}{2n+d}\;\{x^{2n}\}\;\;,
\ee

\noindent with $\{x^{0}\}=1$. For $d\to\infty$ one has the following solution

\be{\label{eq:recapp}}
\{x^{2n}\}=\frac{(2n-1)!!}{d^{n}}\;\;.
\ee

\noindent Of course in actual computations one must use the exact recursion (\ref{eq:recex}). The point we wanted to expose is that this will allow for a $1/d$ expansion of (\ref{eq:gene}) in terms of the even derivatives of $F(\sqrt{1+\lambda^{2}\rho^{2}+2\lambda\rho x})$ with respect to $x$ evaluated at $x=0$. The meaning
of this is clear if we remember the meaning of the measure $d\omega_{d}(x)$ in (\ref{eq:measx}) : as $d\to\infty$ for all $x$
except for $x=0$ this measure represents a number smaller than unity raised to a very large
power. At $x=0$ the measure is $1$. Therefore as $d\to\infty$ the normalized measure $d\omega_{d}(x)/\omega_{d}$ represents a properly normalized $\delta$ function,

\be
\lim_{d\to\infty} \frac{d\omega_{d}(x)}{\omega_{d}}\to \delta(x)dx\;\;.
\ee

We therefor come to the following in which  we use a condensed notation

\be{\label{eq:first}}
\left<F(r)\right>=\sum_{n_{1}=0}^{\infty}\;\frac{\{x^{2n_{1}}\}}{2n_{1}!}\;\left<F^{2n_{1}}(\sqrt{1+\lambda^{2}r^{2}})\right>\;\;,
\ee

\noindent where we have

\begin{subequations}
\bea
F^{2n_{1}}(s_{1}(r,0,\lambda))&=&\frac{d^{2n_{1}}F\left(s_{1}(r,x,\lambda)\right)}{dx^{2n_{1}}}\mid_{x=0}\;,\\
s_{1}(r,x,\lambda)&=&\sqrt{1+\lambda^{2}r^{2}+2\lambda rx}\;\;.
\eea
\end{subequations}

The expression in (\ref{eq:first}) is not a complete expansion in $1/d$ since one still has to evaluate $<F^{2n_{1}}(\sqrt{1+\lambda^{2}r^{2}})>$. But we can repeat the whole procedure between equations (\ref{eq:gene}) and (\ref{eq:first}) many more times. And the resulting form after $k$ iterations is

\bea{\label{eq:kth}}
\left<F(r)\right>=\sum_{n_{1}=0}^{\infty}&\cdots&\sum_{n_{k}=0}^{\infty}g(n_{1},\cdots,n_{k})\nonumber\\
&\times&\left<F^{2n_{1},\cdots,2n_{k}}(s_{k}(r,0,\lambda))\right>\;\;.
\eea

\noindent With the following terms

\be
g(n_{1},\cdots,n_{k})=\prod_{i=1}^{k}\frac{\{x^{2n_{i}}\}}{2n_{i}!}\;\;,
\ee
\be
s_{k}(r,x,\lambda)=\sqrt{\frac{1-\lambda^{2k}}{1-\lambda^{2}}+\lambda^{2k}r^{2}+2\lambda^{2k-1}rx}\;\;,
\ee
\bea
&F&^{2n_{1},\cdots,2n_{k}}(s_{k}(r,0,\lambda))=\nonumber\\
&&\frac{d^{2n_{k}}F^{2n_{1},\cdots,2n_{k-1}}\left(s_{k}(r,x,\lambda)\right)}{dx^{2n_{k}}}\mid_{x=0}\;\;.
\eea

Considering an infinite iteration of the procedure we arrive at the completed $1/d$ expansion

\bea{\label{eq:inftth}}
\left<F(r)\right>=\sum_{n_{1}=0}^{\infty}&\cdots&\sum_{n_{\infty}=0}^{\infty}g(n_{1},\cdots,n_{\infty})\nonumber\\ 
&\times& F^{2n_{1},\cdots,2n_{\infty}}\left(\sqrt{\frac{1}{1-\lambda^{2}}}\right)\;.
\eea

It is instructive to note that in this expression we evaluate $F^{2n_{1},\cdots,2n_{\infty}}$ at the value 
$r=1/\sqrt{1-\lambda^{2}}$ implying that they are evaluated at $d=\infty$. The reason for this is the same as
the large $N$ argument for $O(N)$ field theories: if $N=\infty$ only planar diagrams will contribute to the generator of connected green functions and all the higher correlations factorize in terms of the two-point expectations. The correspondence to our model is that at $d=\infty$, our $d$-dimensional random walker in the infinite time limit, is like a zero-dimensional $O(d)$ field theory. So when $d=\infty$ the only non-trivial moment is $<r^{2}>$ and all the others are given as

\be
\lim_{d\to\infty}\left< r^{p}\right> = \left< r^{2} \right> ^{p/2} = \left(\frac{1}{1-\lambda^{2}}\right)^{p/2}\;,
\ee

So consequently we have

\be
\lim_{d\to\infty}\left<F(r)\right>=F\left(\frac{1}{\sqrt{1-\lambda^{2}}}\right)\;.
\ee

\noindent As it is implicitly assumed, for this expansion to work the $F$ should be a smooth function.

\subsection{Analysis of Lower Cumulants}

We do not know {\em{a priori}} the form of the distribution $P_{\lambda}$ and an exact treatment is not immediately evident. However, one can obtain valuable information for the form of the distribution by studying lower moments or cumulants. Two of the most important quantities are the skewness and the kurtosis of a distribution
which we will study below.

\subsubsection{Lower cumulants in 1 dimensions}
In one dimensions we have the following from previous sections

\begin{subequations}
\bea
\mathcal{S}&=&0\;\;,\\
\mathcal{C}&\equiv&\frac{\left<z^{4}\right>-3\left<z^{2}\right>}{\left<z^{2}\right>^{2}}=-2\frac{1-\lambda^{2}}{1+\lambda^{2}}\;\;.
\eea
\end{subequations}

\noindent Skewness $\mathcal{S}$ vanishes as one would expect for a symmetric distribution. However the kurtosis excess $\mathcal{C}\leq0$. A negative kurtosis excess means that the peak of the distribution is flatter ({\em platykurtic, subgaussian}) than the
gaussian distribution for which $\mathcal{C}=0$. For $\lambda\to1$ we recover a vanishing kurtosis excess  as one would expect, since for this value of $\lambda$ the distribution is a gaussian in the infinite time limit.

\subsubsection{$1/d$ expansion of cumulants in $d$-dimensions}

In $d$-dimensions, in our normalization scheme, one should actually pay attention to $r^{d-1} P_{\lambda}(r)$. This function generally has non-vanishing odd moments $\left<r^{2n+1}\right>$. To obtain the kurtosis one needs $n=1$ and $n=3$ along with our exact results for $\left<r^{2}\right>$ and $\left<r^{4}\right>$.  Below,
we present the results of a $1/d$ analysis up to and including order $1/d^{2}$. The cumulants are defined as,

\begin{subequations}
\bea
c_{2}&=&\left<r^{2}\right>-\left<r\right>^{2}\;,\\
c_{3}&=&2\left<r\right>^{3}-3\left<r\right>\left<r^{2}\right>+\left<r^{3}\right>\;,\\
c_{4}&=&-6\left<r\right>^{4}+12\left<r^{2}\right>\left<r\right>^{2}-3\left<r^{2}\right>^{2}\;,\nonumber\\
&&-4\left<r\right>\left<r^{3}\right>+\left<r^{4}\right>\;.
\eea
\end{subequations}

The final form of cumulants are as follows

\begin{subequations}\label{eq:cumP}
\be
\tilde{c}_{2}=\frac{\lambda^{2}}{d(1+\lambda^{2})}\left[1-\frac{\lambda^{2}(-7+5\lambda^{2}+5\lambda^{4})}{2d(1+\lambda^{2})(1+\lambda^{2}+\lambda^{4})}\right],
\ee
\be
\tilde{c}_{3}=-\frac{1}{d^{2}}\frac{3\lambda^{4}(1-\lambda^{2}-\lambda^{4})}{(1+\lambda^{2})^{2}(1+\lambda^{2}+\lambda^{4})}\;,
\ee
\be
\tilde{c}_{4}=0\;+\mathcal{O}(\frac{1}{d^{3}})\;,
\ee
\end{subequations}

\noindent where we have normalized the cumulants as follows

\be
\tilde{c}_{n}\equiv\frac{c_{n}}{\left<r^{2}\right>^{n/2}}\;.
\ee

We would like to compare these cumulants to that of a gaussian in $d$-dimensions

\be
{\rm G(r)}=N_{d}\;r^{d-1}\;\exp\left(-\frac{r^{2}}{2\sigma^{2}_{d}}\right)\;.
\ee

\noindent Here $N_{d}$ is the normalization factor and $\sigma^{2}_{d}=\sigma^{2}/d$ so that $\left<r^{2}\right>=\sigma^{2}$ regardless of dimensions. Performing the same analysis for ${\rm G}$ we
find the following,

\begin{subequations}{\label{eq:cumG}}
\bea
\tilde{c}^{G}_{2}&=&\frac{1}{2d}\left[1-\frac{1}{4d}\right]\;\;,\\
\tilde{c}^{G}_{3}&=&\frac{1}{4d^{2}}\;\;,\\
\tilde{c}^{G}_{4}&=&0+\frac{5}{16d^{3}}+\mathcal{O}(\frac{1}{d^4})\;\;,
\eea
\end{subequations}

\noindent which are expansions of the results one can obtain in closed form for $G$.

A comparison of cumulants in (\ref{eq:cumP}) and (\ref{eq:cumG}) reveals an exact match for $\lambda=1$ as one should expect. On the other hand, it is also evident that to  order $1/d$, the cumulants of $P_{\lambda}$ have the form of the gaussian, albeit with a scale mismatch for the variance. That is, to order $1/d$ the distribution of the random walker we study is very close to a  gaussian. This can be substantiated if we go back to the master equation (\ref{eq:pdf}) for $P_{\lambda}$ and expand up to order $1/d$

\be{\label{eq:pdf1overd}}
P_{\lambda}(\vec{r},N+1)=P_{\lambda}(\vec{r},N)+\frac{\lambda^{2N}\delta x^{2}}{2d}\nabla^{2}P_{\lambda}(\vec{r})+\cdots\;\;,
\ee

\noindent where we have used 

\[
\left<\hat{r}_{n}^{i}\hat{r}_{n}^{j}\right>=\frac{\delta^{ij}}{d}\;\;,
\]

\noindent and introduced a length scale $\delta x$ via $\hat{r}\to \delta x\hat{r}$. One can also introduce a time scale $\delta t$ so that one will end up with the following form

\be
\frac{\partial P}{\partial t}=\frac{1}{2 d}\;\frac{\delta x^{2}}{\delta t}\; \lambda^{2 t/\delta t} \nabla^{2}P+\cdots\;\;.
\ee

One can now try to do the usual practice of going to the continuum limit by taking $\delta x\to0$ along with $\delta t\to 0$ such that the ratio $\delta x^{2}/\delta t$ is the diffusion constant

\be
\lim_{\delta x\to 0} \lim_{\delta t\to0} \frac{\delta x^{2}}{\delta t}=D\;\;.
\ee

However there is a nuance; the factor $\lambda^{2N}$ with $N=t/\delta t$ spoils this limit.
So one must require the following as well,

\be
\lim_{\delta t\to 0} \lim_{\lambda \to 1} \lambda^{2 t/\delta t}=f(t/\tau).
\ee

\noindent Where the function $f$ depends on how the limit is taken. The most natural choice for $f$ that has all the flavor of our model is the exponential function $f(t/\tau)=\exp(-t/\tau)$. That is we take the limit $\delta t\to0$ along with $\lambda\to 1$ such that

\be
-2 \frac{\delta t}{\ln(\lambda)}=\tau.
\ee

The complications that arise in this {\bf would be} continuum limit are related to the
fact that in truth the random walker with shrinking steps does not walk on a static background of lattice points. The support of the distribution for any discrete time $N$ is a set that evolves in a complicated way everytime the walker takes a step, which prevents  a continuum limit for $\lambda\neq 1$. We therefor arrive at the following $1/d$ {\bf continuum counterpart} system

\be
\frac{\partial{P}}{\partial t}=\frac{D}{2 d}\exp(-t/\tau)\;\nabla^{2}P\;\;.
\ee

This equation can better be written in terms of a different time variable

\be{\label{eq:ttoT}}
T\equiv \tau\left[1-\exp(-t/\tau)\right]\;\;,
\ee

\noindent the limits $t\in(0,\infty)$ corresponding to $T\in(0,1)$. And the equation becomes the common diffusion equation

\be{\label{eq:continuum}}
\frac{\partial P}{\partial T}=\frac{D}{2d}\;\nabla^{2}P\;.
\ee

In \cite{art3}, where the first passage characteristics of the random walker with shrinking steps
were studied, is was shown that the continuum counterpart (\ref{eq:continuum}) captured all the qualitative aspects of the system with the equivalence of parameters $\tau\sim-1/\ln(\lambda)$.

We therefor have established that to $\mathcal{O}(1/d)$ the distribution function $P_{\lambda}$ is like a gaussian with a non-trivial rescaling of the variance, which is
due to the non-triviality of the limit that resulted in the continuum counterpart equation.

There are deviations in the cumulants in (\ref{eq:cumP}) and (\ref{eq:cumG}) starting at $\mathcal{O}(1/d^{2})$. The variance $\tilde{c}_{2}>0$ for all $\lambda$ and there is a critical value of $\lambda_{var}$ which can make the $1/d^{2}$ term to vanish (but $\tilde{c}_{3}$ survives it) 

\[
\lambda_{var} = \sqrt{\frac{\sqrt{33/5}-1}{2}}\;.
\]

The skewness given by $\tilde{c}_{3}$ changes sign when the value of $\lambda$ becomes the square root of golden ratio, 

\[
\lambda_{skew}=\sqrt{\frac{\sqrt{5}-1}{2}}\;\;.
\]

\noindent Which means that to $\mathcal{O}(1/d)$ and for $\lambda=\lambda_{skew}$ the distribution $P_{\lambda}(r)r^{d-1}$ is symmetric around $<r>$.

Higher order expansions are in principle possible and could give further insight for $P_{\lambda}$ but we do not pursue them here.

\section{Long Term Correlations}

In this section we study some of the long term memory effects of the walker at hand. Since the walker ultimately comes to a halt due to the shrinking of the step sizes, it can not be very far away from the position of the first step of unit length, which we can always choose  to be  the $+\hat{z}$ direction in a $d$-dimensional space. Thus the final position will have some memory of this first step unlike
a normal random walker where the memory is washed out in the infinite time limit. 

\subsection{Simple Examples: $\left<z\right>$ and $\left<z^2\right>$}

It is fairly straightforward to study quantities related to lower moments of the distribution. For example what would be the average position as time goes to infinity along the $z$ direction given that the first step was along the positive $z$ direction? Alternatively one could ask the same for the mean square displacement along the $z$ direction. Applying the scaling property we get

\bea
\left<z_{\infty}\right>&=&1\;\;,\\
\left<z_{\infty}^{2}\right>&=&1+\frac{1}{d}\left(\frac{\lambda^{2}}{1-\lambda^{2}}\right)\;\;.
\eea

These are fairly easy to understand. The mean $\left<z_{\infty}\right>$ would be $1$ even for the ordinary random walk. But the fact that our walker eventually stops changes the behavior of $\left<z_{\infty}^{2}\right>$ which is expected to diverge as time goes to infinity, in the case of the ordinary random walk. The fact that $\left<z_{\infty}^{2}\right>\to 1$ as $d\to\infty$  is also expected: as the number of dimensionality grows the possibility to walk along a single direction decreases. In fact as $d\to\infty$ all the subsequent walk following the first step can, to a very good accuracy, be considered to be happening along the directions orthogonal to $\hat{z}$.

\subsection{The correlation $\left<\cos(\theta)\right>$}

Our walker will not be infinitely far away from its first
step. This, in turn, will introduce an angle correlation between the first step which is at $\hat{z}$ and the last step $\vec{r}_{\infty}$. An interesting quantity to study is the cosine of this angle,

\be
\left<\cos(\theta)\right>=\left<\frac{\vec{r}_{1}\cdot\vec{r}}{|\vec{r}_{1}||\vec{r}|}\right>\;\;.
\ee

For the ordinary random walker this correlation should vanish in the infinite time limit since the walker would have spread all over the space equally likely. In this subsection we will analyze this
angle distribution for one and three dimensional cases and finally present a $1/d$ expansion.

\subsubsection{One dimensional walker}

In one dimensions \footnote{In one dimensions, we denote the position by $z$ in accord with our formalism and normalize again via integration along positive $z$ only.} the correlation we are after simplifies to

\be
\left<\cos(\theta)\right>=\left<{\rm Sign}(z)\right>
\ee

\noindent That is, we would be computing, in the infinite time limit, the probability that the particle is at $z>0$ minus the probability that it is at $z<0$ given that the first step is to the right. This will be related to the ultimate survival (remaining at $z>0$) probability of a walker if the first step was to the right.

It is not difficult to arrive at the following result after a direct implementation of the scaling property

\be
\left<{\rm Sign}(z)\right>=\int_{0}^{1/\lambda}\;P_{\lambda}(z)dz
\ee

At this point we would like to remind an important point about the limits of integration in general. Since the maximum distance the walker can travel is $1/(1-\lambda)$ the distribution will vanish for greater displacements . For $\lambda\leq1/2$ we have the following situation

\be
\lambda\leq\frac{1}{2} \Longrightarrow \frac{1}{\lambda}\geq\frac{1}{1-\lambda}\equiv z_{max}
\ee

\noindent which in turn means that

\be
\lambda\leq\frac{1}{2} \Longrightarrow \left<{\rm Sign}(z)\right>=1
\ee

\noindent Thus for $\lambda\leq1/2$ the cosine of the angle between the first step and the last position are fully correlated, the reason is that the walker will fall out of range of the origin once it takes a step to the right.

On the other hand as $\lambda\to1$ we have $P_{\lambda}\to 0$ and hence the correlation vanishes. The walker cannot remember it took the first step to the right as one should expect.

\begin{figure}[t]
\includegraphics[scale=0.35,angle=270]{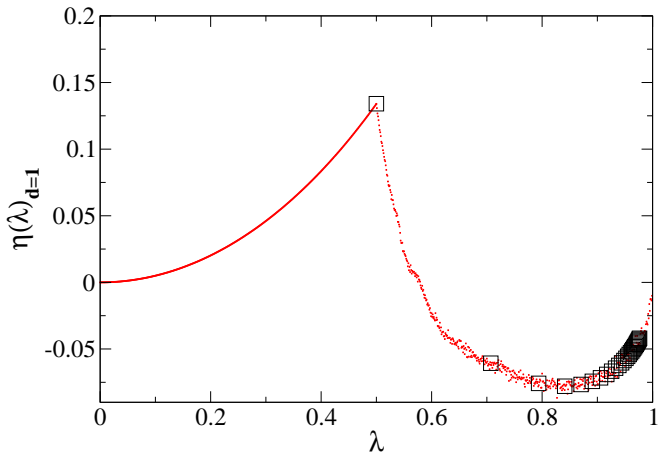}
\caption{{\label{fig:a2}} Plot of $\eta(\lambda)_{d}$ for $d=1$, defined in (\ref{eq:shiftedcos}). The background points are results of a numerical simulation. The squares represent analytical results for $\lambda=2^{-1/m}$, with $m$ from 1 to 30.}
\end{figure}

The results of a numerical  study of the problem are represented in Fig.~\ref{fig:a2}. We have also attempted an analytical study of for $\lambda=2^{-1/m}$. It has been shown in  \cite{art1} and \cite{art2} that for these values of $\lambda$ the distribution $P_{\lambda}$ are $m$'th order convolutions of a uniform distribution and hence
a direct analytical evaluation is possible, albeit via symbolic programming packages. The squares in Fig.~\ref{fig:a2} denote results for $m\in \mathcal{N}$.

\subsubsection{Three dimensional walker}

For a three dimensional walker we have to evaluate the following 

\be
\left<\cos(\theta)\right>=\frac{1}{2}\int \;dx\;du\;u^{2} \frac{1+\lambda u x}{\left|\hat{z}+\lambda\vec{u}\right|}\;P_{\lambda}(u)
\ee

The denominator can be expanded using the Legendre polynomials

\be
\frac{1}{\left|\hat{z}+\lambda\vec{u}\right|}=\sum_{l=0} (-1)^{l}p_{l}(x)\left(\frac{u_{<}}{u_{>}}\right)\;\frac{1}{u_{>}}
\ee

\noindent with $u_{>}$ and $u_{<}$ are either $1$ or $\lambda u$ depending on which one is bigger.

\vspace*{1cm}

\begin{figure}[t]
\includegraphics[scale=0.35,angle=270]{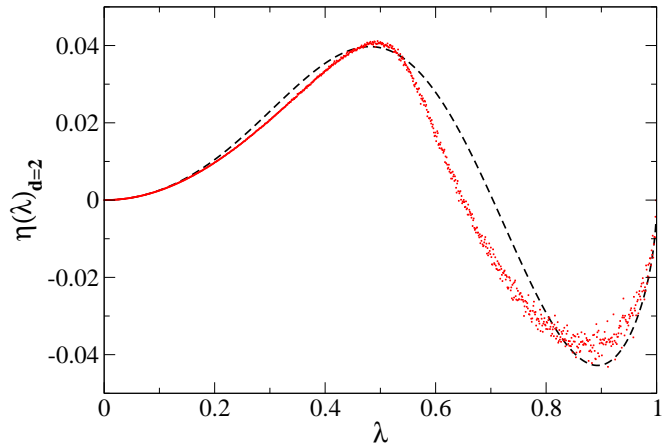}
\caption{{\label{fig:a3}}Plot of $\eta(\lambda)_{d}$ for $d=2$. The dashed line represents the $1/d$ expansion in the text. The other plot is the result of a numerical simulation.}
\end{figure}

There is again a simplification if $\lambda\leq1/2$. In this case when $u_{<}=\lambda u$ the upper limit of integration on $u$ is $1/\lambda$ which is greater than the maximum range. So in this case any function of $u$ under the integrand  can be seen as an expectation value. There is a further simplification for $d=3$. Since
the angular measure is just $dx$ we can use the orthogonality properties of the Legendre polynomials. This results in 

\be{\label{eq:feeble}}
\lambda\leq\frac{1}{2}\;{\rm for }\;d=3 \Longrightarrow \;\left<\cos(\theta)\right>=\frac{3-4\lambda^{2}}{3(1-\lambda^{2})}\;\;.
\ee

Unfortunately such a simplification is not present for $\lambda\geq1/2$. We present a computer simulation
of $\left<\cos(\theta)\right>$ in Fig.~\ref{fig:a3} along with the result in (\ref{eq:feeble}).

\begin{figure}[t]
\includegraphics[scale=0.35,angle=270]{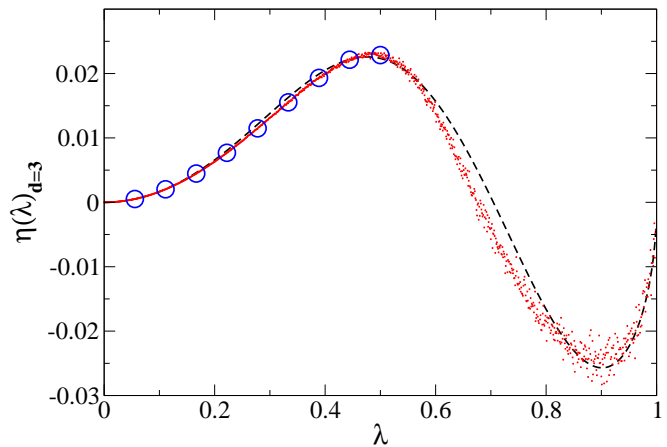}
\caption{{\label{fig:a4}}Plot of $\eta(\lambda)_{d}$ for $d=3$. The dashed line represents the $1/d$ expansion in the text. The circles are representative points for the exact result (\ref{eq:feeble})obtained for $\lambda<1/2$. The other plot is the result of a numerical simulation.}
\end{figure}\vspace*{1cm}

\begin{figure}[t]
\includegraphics[scale=0.35,angle=270]{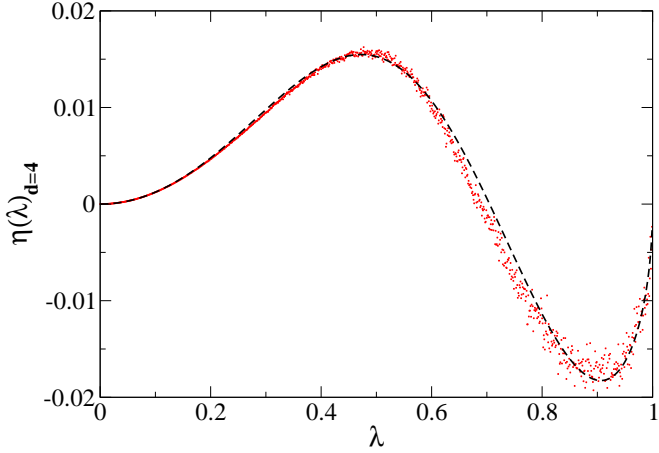}
\caption{{\label{fig:a5}} Plot of $\eta(\lambda)_{d}$ for $d=4$. The dashed line represents the $1/d$ expansion in the text. The other plot is the result of a numerical simulation.}
\end{figure}

\subsubsection{$d$-dimensional walker}

In general dimensions one can still make use of the Legendre polynomial expansion, but this gets complicated rather fast and allows for an exact treatment only if $\lambda<1/2$. One can, on the other hand, use the $1/d$ expansion formalism we presented. Some algebra yields the following up to $1/d^{2}$

\be{\label{eq:1overdexp}}
\left<\cos(\theta)\right>=\sqrt{1-\lambda^{2}}\left[1+a(\lambda)\frac{1}{d}+b(\lambda)\frac{1}{d^{2}}+\cdots\;\right]\;\;.
\ee

\noindent The first term is the $d=\infty$ value. As can be easily checked $\left<\cos(\theta)\right>=<z/r>$ and since in this limit correlations factorize we can use $<z>=1$ and $r^{-1}=\sqrt{r^{-2}}=\sqrt{1-\lambda^{2}}$. The other terms are 

\begin{subequations}
\be
a(\lambda)=\frac{\lambda^{2}(1-2\lambda^{2})}{2(1+\lambda^{2})}
\ee
\bea
b(\lambda)=&&\frac{3\lambda^{4}}{8}\;\left(16\lambda^{8}+12\lambda^{6}-13\lambda^{4}-29\lambda^{2}+15\right)\nonumber\\
&\times&\left((1+\lambda^{2})^{2}(1+\lambda^{2}+\lambda^{4})\right)^{-1}\;.
\eea
\end{subequations}

\noindent Higher order expansion is possible but we do not pursue it here. The series converges rapidly for this observable. We demonstrate this fact by plotting the following quantity both for a set of numerical simulation and the $1/d$ expansion

\be{\label{eq:shiftedcos}}
\eta(\lambda)_{d}\equiv\left<\cos(\theta)\right>_{d}-\left<\cos(\theta)\right>_{d=\infty}\;\;.
\ee

In $d=2$ the agreement is marginal and one has to proceed  to a higher order expansion here. For $d=3$ the agreement is good and already from $d=4$ we have considerable convergence. 

As one can see $\eta(\lambda)$ changes sign meaning that at this particular value of lambda the observable $\left<\cos\theta\right>$ is given by its $d=\infty$ limit. These critical values can be found from our $1/d$ expansion and they are given as follows

\be
\lambda_{c}\approx \frac{1}{\sqrt{2}}\;\left[1-\frac{1}{56 d}+\cdots\right]\;\;.
\ee

\subsection{Ultimate Fate to be in $z>0$}


Another interesting quantity is the probability $p^{+}$, as time goes to infinity, to be at positive $z$ values given that the walker took the first step along the $+\hat{z}$ direction. The reciprocal quantity $p^{-}$ is to be at $z<0$. These quantities are related by the conservation of probability, 

\begin{subequations}
\bea
p^{+}+p^{-}&=&1\\
p^{+}-p^{-}&=&\left<\rm{Sign}(z)\right>
\eea
\end{subequations}

\noindent It is clear, due to the non-differentiability of the signum function, that the $1/d$ expansion procedure will not be directly applicable as in the previous section. Nevertheless the behavior at $d=\infty$ is clear: the walker should remain at the positive $z$ part of space. The large number of dimensions makes it impossible to choose many steps towards a particular direction. This is also immediate from the scale invariance of expectation values: we have $\left<\rm{Sign}(z)\right>=\left<\{\rm{Sign}(1+\lambda r x)\right\}>$ and $d=\infty$ means $x=0$ resulting in $p^{+}-p^{-}=1$ and hence $p^{-}=0$. We also expect, as $\lambda\to 1$, that $p^{+}=p^{-}=1/2$: the usual random walker spreads evenly in the infinite time limit no matter what the initial position is. Moreover, we again have a simplification for $\lambda\leq1/2$: if the walker takes the first step in the $+\hat{z}$ direction the origin and hence the negative $z$ values are out of range. Thus, for $\lambda\leq 1/2$ we have $p^{-}=0$.

The difficulty of a direct computation is revealed if we find the integral expression for $p^{-}$,

\be
p^{-}=\int_{1/\lambda}^{r_{max}}dr\;r^{d-1}\;P_{\lambda}(r)\;\int_{-1}^{-1/\lambda r}\;\frac{d\omega_{d}(x)}{\omega_{d}}\;\;.
\ee

\noindent Since the region of integration on $r$ do not cover the full support of the distribution we can no longer apply our $1/d$ program. Nevertheless the integral over $x$ can be carried out exactly and this results in the following,

\be{\label{eq:survfin}}
2p^{-}=\int_{1/\lambda}^{r_{max}}drr^{d-1}P_{\lambda}(r)\left[1- \mathcal{B}\left(\frac{1}{\lambda^{2}r^{2}},\frac{1}{2},\frac{d-1}{2}\right)\right]\;\;.
\ee

\noindent Here $\mathcal{B}(x,a,b)$ is the regularized incomplete beta function.

We can gain further insight into the $d$ behavior of $p^{+}$ if we are content with the continuum counterpart model we presented in (\ref{eq:continuum}) for which the solution is simply the 
solution of an ordinary random walker with $t$ replaced with $T$ in (\ref{eq:ttoT}). In this scheme we can assume that the walker initially starts at $\vec{r}_{o}=z_{o}\hat{z}$ and find the probability to be at positive $z$ values in the infinite time limit. A simple computation yields,

\be{\label{eq:survcont}}
p_{cont}^{+}=\frac{1}{2}\left[1+{\rm Erf}\left(\sqrt{\frac{d z_{o}^{2}}{2 D \tau}}\right)\right]\;\;,
\ee

\begin{figure}[t]\label{fig:surv}
\includegraphics[scale=0.35,angle=270]{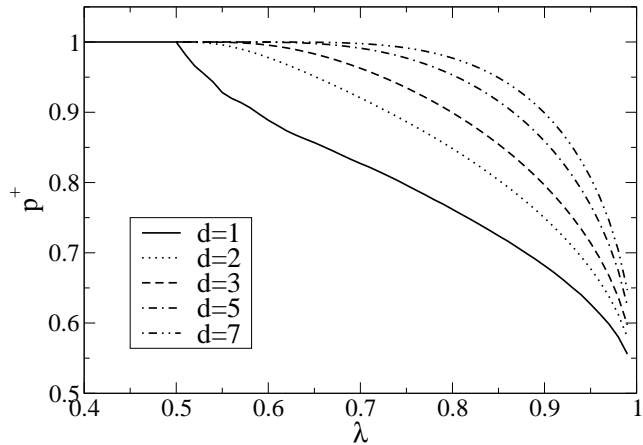}
\caption{{\label{fig:a6}} Simulation results of $p^{+}$ for various dimensions. $\lambda$ range is upto $0.99$. }
\end{figure}

\vspace*{1cm}

\begin{figure}[t]{\label{tab:1}}
\includegraphics[scale=0.35,angle=270]{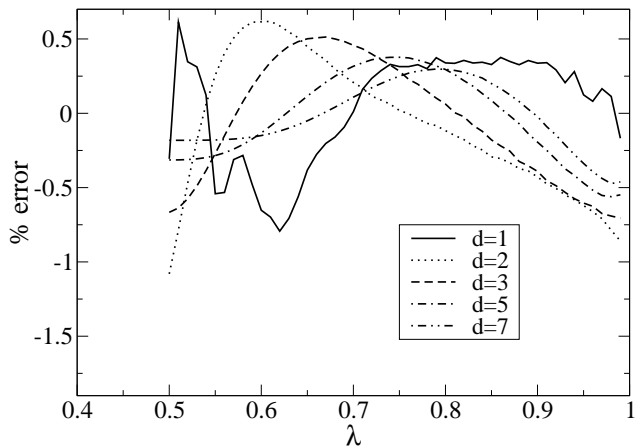}
\caption{{\label{fig:a7}} Percent error between the simulation data and the fit function in (\ref{eq:survfit}) with parameters given in TABLE~I.}
\end{figure}

\noindent where $\rm Erf$ is the error function. Apparently the dependence on $d$ is not what one could recover via a $1/d$ expansion. However, let us  remember, from previous sections, that to order $1/d$, $P_{\lambda}$ is very close to its continuum counterpart system. This, hints at the possibility that the continuum counterpart result could be a very good approximation for any $d$. This proposition can be expanded a bit better as follows. If one wants to compute $p^{+}$ one can just as well proceed first by computing the probability distribution to be at $(z,z+dz)$ by integrating the other dimensions out. This will result in a one dimensional distribution which will be the same as the distribution of a one dimensional walker taking steps of length $\lambda^{n} \cos(\theta_{n})$ at each time step $n$, with $\cos(\theta_{n})$ distribution given by $d\omega_{d}(x)$. Every observable is directly related to the random walker with shrinking steps in one dimensions, for example one will have $<z^{2}>_{d}=<x^{2}>_{d=1}/d$. Thus
most of the $d$ and $\lambda$ dependence might come from (\ref{eq:survcont}). Pursuing this observation and the parametric analogy between $\tau$ and $\lambda$

\[
\tau \sim -\frac{1}{\ln(\lambda)} \; ,
\]

\noindent we can follow a more phenomenological approach. We can try to fit to the following estimate function 

\be{\label{eq:survfit}}
p_{estm}^{+}= \frac{1}{2} A(d)\;\left[1+{\rm Erf}\left(\frac{B(d)}{\sqrt{2}}\sqrt{-d \ln(2 \lambda-1)}\right)\right]\;\;,
\ee

\noindent which has all the global properties of $p^{+}$ mentioned at the beginning of this section.

In TABLE~I we present the results of fits of $p^{+}$ obtained from numerical simulations to the function in (\ref{eq:survfit}). $\chi^{2}$ represents the chi-squared value of the fit and $c$ is its correlation factor. As can be seen the fit is generally good throughout the range of $d$. As expected the fit has an increasing accuracy with increasing $d$.

In Fig.~\ref{fig:a6} one can find the simulation results for $p^{+}$ for various dimensionalities. In Fig~
\ref{fig:a7} we show the percentage error ,as a function of $\lambda$, between the simulation results and the fit function with parameters given in TABLE~I. As can be seen the agreement is good for all values of $\lambda$.

\vspace*{0.5cm}

\begin{table}[h]{\label{tabl1}}
\begin{center}
\begin{tabular}{|c|c|c|c|c|}
\hline
\;\;d\;\; & \;\;A(d)\;\; & \;\;B(d)\;\; & \;\;$\chi^{2}$ \;\;& \;\;c\;\; \\ \hline\hline
\;\;1 \;\;& 1.00305 & 0.974315 & 4.876\;$10^{-4}$ & 0.999674 \\ \hline
\;\;2\;\; & 1.01298 & 0.976132 & 6.222\;$10^{-4}$ & 0.999708 \\ \hline
\;\;3\;\; & 1.00667 & 1.003400 & 6.491\;$10^{-4}$ & 0.999596 \\ \hline
\;\;5\;\; & 1.00330 & 1.015200 & 3.378\;$10^{-4}$ & 0.999702 \\ \hline
\;\;7 \;\;& 1.00183 & 1.01426 & 1.747\;$10^{-4}$  & 0.999800 \\ \hline
\;\;9\;\; & 1.00107 & 1.01129 & 8.787\;$10^{-5}$ &  0.999875 \\ \hline
\end{tabular}
\end{center}
\caption{Fit parameters of the simulation data to the estimate function in (\ref{eq:survfit}).}
\end{table}

\section{Conclusions and Future Directions}

We have tried to expose some of the important properties of a random walker with step sizes shrinking geometrically with each step. The most useful property in general is the scaling property of the probability distribution when one increments the time step by one. This allowed for a $1/d$ expansion approach the same way it would for the ordinary random walker. However, in the model studied this approach remains useful in the infinite time limit where the distribution goes to a time independent function.

We have presented some interesting correlations of the position at large times to the first step. To that end we showed one correlation with a nice $1/d$ expansion and one without. The continuum counterpart system is shown to capture all the qualitative aspects of the system and in general it proves to be a good starting point to approximate the quantitative properties as well.

As in most statistical systems one, two and three dimensional systems are more interesting than higher dimensions where the $1/d$ approach -actually reminiscent of a mean field approach- becomes more and more reliable. One hint for this is already apparent in FIG~5 and TABLE~I, where the curves and fit parameters are characteristically different than the ones for higher dimensionalities. The common topological differences between two and higher dimensions remain. Consequently, for low dimensions, a more incisive approach to the system possibly including the direct study of the probability distribution is important and is left for  future work.

Another possibly interesting avenue to explore is to use lattice schemes to pick hopping directions. Similarly, this should matter only for low dimensions since the mean field should set in regardless for higher dimensions.

We have also left out the first passage properties of the system at hand since it requires a study of the distribution for finite times. This could also prove to be of interest as a topic for further research. The continuum counterpart system should prove to be a nice starting point, for example, for the survival probability in $z>0$.

\acknowledgements{}
I would like to thank M. Mungan for various stimulating discussions and suggestions to enlarge the manuscript. I have also benefited from our discussions and previous collaboration with S. Taneri. I am also grateful to K. Fortuny for the careful reading of the manuscript. The simulations required for this work have been performed on Gilgamesh, a 16-node computer cluster of the Feza G{\"u}rsey Institute.


\begin{thebibliography}{20}
\bibitem{art1} A.C. de la Torre, A. Maltz, H.O. Martin, P. Catuogno and L. Garcia-Mata, Random walk with an exponentially varying step, Phys. Rev. E{\bf{62}}, 7748, 2000.
\bibitem{art2} P.L. Krapivsky and S. Redner, Random walk with shrinking steps, Am. J. Phys. {\bf{72}}(5), 591,  2004
\bibitem{art3} T. Rador and S. Taneri, Random walks with shrinking steps: first-passage characteristics, Phys. Rev. {\bf{E}73}, 036118, 2006.
\bibitem{art4} A.N. Berker and S. Ostlund, J. Phys. C{\bf 12}, 4961, 1979.
\bibitem{art5} T. Niemeijer and J.M.J. van Leeuwen, in: Phase transitions and critical phenomena, Vol. 6., eds. C. Domb and M.S. Green (Academic Press, New York, 1976). 
\bibitem{art6} C. Itzykson and J.M. Drouffe, Statistical Field Theory, Vol. 1., Ch. 4, page 170, (Cambridge Monographs on Mathematical Physics).
\bibitem{art7} A. Erzan, Finite q-differences and the discrete renormalization group, Phys. Lett. A{\bf 225}, 235, 1997; A. Erzan, J.P Eckmann, q-analysis of fractal sets, Phys. Rev. Lett. {\bf{78}}(17), 3245, 1997.
\bibitem{art8} M. Ar{\i}k and D.D. Coon, J. Math. Phys. 17, 524, 1975; M. Ar{\i}k, D.D. Coon and Y.M. Lam, ibid. 16, 1776, 1975; M. Ar{\i}k, Z. Phys. C 51, 627, 1991.
\bibitem{art9} F.H. Jackson, Q. J. Pure Appl. Math. 41, 193, 1910; F.H. Jackson, Q. J. Math. Oxford Ser. 2, 1, 1951.
\end{thebibliography}
\end{document}